\documentclass[12pt,showpacs,aps,prd]{revtex4}
\usepackage{graphicx}
\input epsf

\begin{document}

\title{Can $X(3872)$ be a $J^{P}=2^{-}$ tetraquark state?}
\author{Chun-Yu Cui, Yong-Lu Liu, Guo-Bin Zhang and Ming-Qiu Huang}
\affiliation{Department of Physics, National University of Defense
Technology, Hunan 410073, China}

\date{\today}
\begin{abstract}
In this article, we test the nature of $X(3872)$, which is assumed to be a P-wave $[cq]$-scalar-diquark $[\bar{c}\bar{q}]$-axial-vector-antidiquark tetraquark state with $J^{P}=2^{-}$. The interpolating current representing the $J^{P}=2^{-}$ state is proposed. Technically, contributions of the operators up to dimension six are included in the operator product expansion (OPE). The mass obtained for such state is  $m_{2^{-}}=(4.38\pm 0.15)~\mbox{GeV}$. We conclude that it is impossible to describe the $X(3872)$ structure as $J^{P}=2^{-}$ tetraquark state.
\end{abstract}
\maketitle

The state $X(3872)$ was first discovered by Belle~\cite{Xobservation} in the $\pi^+\pi^-J/\psi$ mode and then confirmed by the CDF~\cite{CDF}, D$\emptyset$~\cite{D0}, and BABAR~\cite{BABAR} Collaborations in the same decay channal. The most recent measure of its mass is~\cite{Belle2}
\begin{equation}
m_X=(3871.85\pm 0.27({\rm stat})\pm0.19({\rm syst}))~\mbox{MeV},
\label{Xmass}
\end{equation}
with a width of $\Gamma_X<1.2~\mbox{MeV}$. Belle~\cite{Xobservation} and CDF~\cite{Abulencia:2006ma} propose that it proceeds through the $X \to J/\psi \rho \to J/\psi \pi^+\pi^-$ decay. Since a charmonium state has isospin zero, it can not decay into $X \to J/\psi \rho $, so the $X(3872)$ is identified as an ``exotic" state. According to the CDF analysis of the decay angular distribution~\cite{Abulencia:2006ma} and the invariant $\pi^+\pi^-$ mass distribution~\cite{Abulencia:2005zc} of the $J/\psi \pi^+ \pi^-$ decay mode, only $1^{+}$ and $2^{-}$ assignments are possible.  The close proximity of $X(3872)$ mass to the $D\bar D^{*}$ threshold indicates that $X(3872)$ might be a loosely bound $D\bar D^{*}$ molecular state, whose quantum number is $J^{P}=1^{+}$. Also, an angular analysis applied to the $2\pi$ mass distribution in $J/\psi\rho$ favors the quantum number $J^P=1^+$~\cite{Belle-X-gamma-psi}. In compliance with these quantum numbers, many literatures have appeared in the past years. Its possible interpretations include the molecular state, tetraquark state and hybrid charmonium (see reviews \cite{Swanson:2006st}-\cite{Nora} and references therein). Using QCD sum rules (QCDSR)~\cite{svzsum}, Nielsen {\it et al.} discuss the possibility that it is possible to describe the $X(3872)$ structure as a mixed molecule-charmonium state and study its strong decay and radiative decay~\cite{Nielsenmix,Nielsenradia}.

Very recently, the BABAR collaboration has performed angular distribution analysis of the decay $B\to J/\psi\omega K$, indicating that P-wave between $J/\psi$ and $\omega$ is favored, so that quantum numbers $J^{P}=2^{-}$ is preferred~\cite{BABAR2}. In this case, the most conventional explanation is the $1^1D_2$ charmonium state $\eta_{c2}(1D)$. In Ref~\cite{Yu1007}, the radiative transition processes $\eta_{c2}(1D)\to J/\psi(\psi')+\gamma$ is investigated within several phenomenological potential models with the assumption that $X(3872)$ is a $\eta_{c2}(1D)$ charmonium, which are in contradiction with the existing BABAR measurements~\cite{CDF2}. The data on its $D^0 \bar{D}^0 \pi^0$ decay mode~\cite{Belle3} also contradict the $1^1D_2$ charmonium interpretation of the $X(3872)$~\cite{Kalash1008}. The decay of $B\to \eta_{c2}X $ is studied in NRQCD factorization framework, which indicates that X(3872) is unlikely to be a $1^1D_2$ charmonium state~\cite{Chao}. Thus, we have to resort to exotic explanations for the $J^{P} = 2^{-}$ quantum numbers. In Ref.~\cite{Polosa}, it's shown that the molecular interpretation appears to be untenable, but the tetraquark interpretation may be a viable candidates to be $X(3872)$ with $J^{P} = 2^{-}$. Follow their opinion, we study the mass of $X(3872)$ as a P-wave $[cq]$-scalar-diquark $[\bar{c}\bar{q}]$-axial-vector-antidiquark tetraquark state with $J^{P}=2^{-}$ using the QCDSR.

The interpolating current representing a $J^{P}=2^{-}$ P-wave tetraquark state with $[cq]$-scalar-diquark and $[\bar{c}\bar{q}]$-axial-vector-antidiquark fields is adopted as
\begin{eqnarray}
j_{\mu\nu}={\epsilon_{abc}\epsilon_{dec}\over\sqrt{2}}[(q_a^T C\gamma_5
c_b)D_{\mu}(\bar{q}_d\gamma_\nu C\bar{c}_e^T)]\;. \label{field2}
\end{eqnarray}
Herein the index $T$ represents matrix transposition, $C$ means the charge conjugation matrix, $D^{\mu}$ denotes the covariant derivative, while $a$, $b$, $c$, $d$, and $e$ are color indices.

In the QCDSR approach, the mass of the particle can be determined by considering the two-point correlation function
\begin{equation}\label{cor}
\Pi_{\mu\nu,\alpha\beta}(q^{2})=i\int d^4xe^{iq.x}<0|T[j_{\mu\nu}(x)j^{+}_{\alpha\beta}(0)]|0>.
\end{equation}
The QCDSR attempts to link the hadron phenomenology with the interactions of quarks and gluons, which is obtained by evaluating the correlation function in two ways: an approximate description of the correlation function in terms of intermediate states through the dispersion relation, a description of the same correlation function in terms of QCD degrees of freedom via OPE.

In the phenomenological side, the correlation function is calculated by inserting a complete set of intermediate states with the same quantum numbers as the tetraquark state. Parametrizing the coupling of the $J^{P}=2^{-}$ tensor state to the current $j _{\mu\nu}$ in term of the parameter $f_{X}$ as
\begin{eqnarray}\label{lep}
\langle 0| j_{\mu\nu}(0)| X \rangle=f_{X} \varepsilon_{\mu\nu},
\end{eqnarray}
where $\varepsilon_{\mu\nu}$ is the relevant polarization tensor. Using Eq. (\ref{lep}) in the phenomenological side of Eq. (\ref{cor}), we obtain
\begin{eqnarray}\label{phen2}
\Pi_{\mu\nu,\alpha\beta}=\frac{f^2_{X}}{m_{X}^2-q^2}
\left\{\frac{1}{2}(g_{\mu\alpha}g_{\nu\beta}+g_{\mu\beta}g_{\nu\alpha})\right\}+
\mbox{other structures}+...,
\end{eqnarray}
where the only structure which contains the contribution of the tensor meson has been kept. In calculations, we have performed summation over the polarization tensor using
\begin{eqnarray}\label{polarizationt1}
\varepsilon_{\mu\nu}\varepsilon_{\alpha\beta}^*=\frac{1}{2}T_{\mu\alpha}T_{\nu\beta}+
\frac{1}{2}T_{\mu\beta}T_{\nu\alpha}
-\frac{1}{3}T_{\mu\nu}T_{\alpha\beta},
\end{eqnarray}
with
\begin{eqnarray}\label{polarizationt2}
T_{\mu\nu}=-g_{\mu\nu}+\frac{q_\mu q_\nu}{m_{X}^2}.
\end{eqnarray}

In the OPE side, we single out the structure $\frac{1}{2}(g_{\mu\alpha}g_{\nu\beta}+g_{\mu\beta}g_{\nu\alpha})$ whose coefficient is denoted as
\begin{eqnarray}\label{ope}
\Pi^{(1)}(q^{2})=\int_{4m_{c}^{2}}^{\infty}ds\frac{\rho^{OPE}(s)}{s-q^{2}},
\end{eqnarray}
where the spectral density is $\rho^{OPE}(s)=\frac{1}{\pi}\mbox{Im}\Pi^{\mbox{(1)}}(s)$. After equating the two sides, assuming quark-hadron duality, and making a Borel transformation, the sum rule can be written as
\begin{eqnarray}\label{sr}
f_{X}^{2}e^{-m_{X}^{2}/M^{2}}&=&\int_{4m_{c}^{2}}^{s_{0}}ds\rho^{OPE}(s)e^{-s/M^{2}},
\end{eqnarray}
with $M^2$ the Borel parameter.

In calculations, we work at the leading order in $\alpha_{s}$ and consider vacuum condensates up to dimension six, with the similar techniques in Refs.~\cite{technique,technique1}. After tedious calculation, the concrete forms of spectral densities read:
\begin{eqnarray}
\rho^{OPE}(s)=\rho^{\mbox{pert}}(s)+\rho^{\langle\bar{q}q\rangle}(s)+\rho^{\langle
g^{2}G^{2}\rangle}(s)+\rho^{\langle
g\bar{q}\sigma\cdot G q\rangle}(s)+\rho^{\langle\bar{q}q\rangle^{2}}(s),
\end{eqnarray}
with
\begin{eqnarray}\label{spectralmole}
\rho^{\mbox{pert}}(s)&=&\frac{1}{5*2^{13}\pi^{6}}\int_{\alpha_{min}}^{\alpha_{max}}\frac{d\alpha}{\alpha^{4}}\int_{\beta_{min}}^{1-\alpha}\frac{d\beta}{\beta^{4}}(\alpha^3+\alpha^2\beta-2\alpha^2+\alpha\beta^2-3\alpha\beta+2\alpha+\beta^3
\nonumber\\&&{}-\beta^2+\beta-1)r(m_{c},s)^{5}
,\nonumber\\
\rho^{\langle\bar{q}q\rangle}(s)&=&\frac{\langle\bar{q}q\rangle}{3*2^{8}\pi^{4}}m_{c}\int_{\alpha_{min}}^{\alpha_{max}}\frac{d\alpha}{\alpha^{2}}\int_{\beta_{min}}^{1-\alpha}\frac{d\beta}{\beta^{2}}(2\beta^2+2\alpha\beta-2\beta-1)r(m_{c},s)^{3}
,\nonumber\\
\rho^{\langle g^{2}G^{2}\rangle}(s)&=&\frac{\langle
g^{2}G^{2}\rangle}{3^{3}*2^{13}\pi^{6}}m_{c}^{2}\int_{\alpha_{min}}^{\alpha_{max}}\frac{d\alpha}{\alpha^{4}}\int_{\beta_{min}}^{1-\alpha}\frac{d\beta}{\beta}(\alpha+\beta-1)^{2}(2\alpha^2+\alpha-2\beta^2-\beta)r(m_{c},s)^{2}
,\nonumber\\
\rho^{\langle g\bar{q}\sigma\cdot G q\rangle}(s)&=&\frac{\langle
g\bar{q}\sigma\cdot G
q\rangle}{2^{10}\pi^{4}}m_{c}\int_{\alpha_{min}}^{\alpha_{max}}\frac{d\alpha}{\alpha(\alpha-1)}[m_{c}^{2}-\alpha(1-\alpha)s]^2
\nonumber\\&&{}
-\frac{\langle g\bar{q}\sigma\cdot
Gq\rangle}{2^{10}\pi^{4}}m_{c}\int_{\alpha_{min}}^{\alpha_{max}}\frac{d\alpha}{\alpha^{2}}\int_{\beta_{min}}^{1-\alpha}\frac{d\beta}{\beta^{2}}(4\alpha^2\beta+6\alpha\beta^2-2\alpha\beta-\alpha^2-\beta)r(m_{c},s)^{2}
,\nonumber\\&&{}
-\frac{\langle g\bar{q}\sigma\cdot
Gq\rangle}{2^{10}\pi^{4}}m_{c}\int_{\alpha_{min}}^{\alpha_{max}}\frac{d\alpha}{\alpha}\int_{\beta_{min}}^{1-\alpha}\frac{d\beta}{\beta^2}(2\beta^2+2\alpha\beta-2\beta-\alpha)r(m_{c},s)^{2}
,\nonumber\\
\rho^{\langle\bar{q}q\rangle^{2}}(s)&=&\frac{\langle\bar{q}q\rangle^{2}}{3*2^{5}\pi^{2}}m_{c}^{2}(\frac{2m_{c}^{2}}{3}-\frac{s}{6})\sqrt{1-4m_{c}^{2}/s},
\end{eqnarray}
with $r(m_{c},s)=(\alpha+\beta)m_{c}^2-\alpha\beta s$. The integration limits are given by $\alpha_{min}=\Big(1-\sqrt{1-4m_{c}^{2}/s}\Big)/2$, $\alpha_{max}=\Big(1+\sqrt{1-4m_{c}^{2}/s}\Big)/2$, and $\beta_{min}=\alpha m_{c}^{2}/(s\alpha-m_{c}^{2})$.

To extract the mass $m_{X}$, we take the derivative of Eq.(\ref{sr}) with respect to $\frac{1}{M^2}$ and then divide the result by itself
\begin{eqnarray}\label{sumrule}
m_{X}^{2}&=&\int_{4m_{c}^{2}}^{s_{0}}ds\rho^{OPE}(s)s
e^{-s/M^{2}}/
\int_{4m_{c}^{2}}^{s_{0}}ds\rho^{OPE}(s)e^{-s/M^{2}}.
\end{eqnarray}

Before the numerical analysis of Eq.(\ref{sumrule}), we first specify the input parameters. The quark mass is taken as $m_{c}=1.23~\mbox{GeV}$ \cite{PDG}. The condensates are $\langle\bar{q}q\rangle=-(0.23)^{3}~\mbox{GeV}^{3}$, $\langle g\bar{q}\sigma\cdot G q\rangle=m_{0}^{2}~\langle\bar{q}q\rangle$, $m_{0}^{2}=0.8~\mbox{GeV}^{2}$, and $\langle g^{2}G^{2}\rangle=0.88~\mbox{GeV}^{4}$~\cite{svzsum}. Complying with the standard procedure of the QCDSR, the threshold $s_{0}$ and Borel parameter $M^{2}$ are varied to find the optimal stability window. There are two criteria (pole dominance and convergence of the OPE) for choosing the Borel parameter $M^{2}$ and threshold $s_{0}$. In general, the continuum threshold $s_{0}$ is a parameter of the calculation which is connected to the mass of the studied state, by the relation $\sqrt{s_0}\approx (m_{X}+0.5\,\mbox{GeV})$.

\begin{figure}
\centering
\includegraphics[totalheight=5cm,width=7cm]{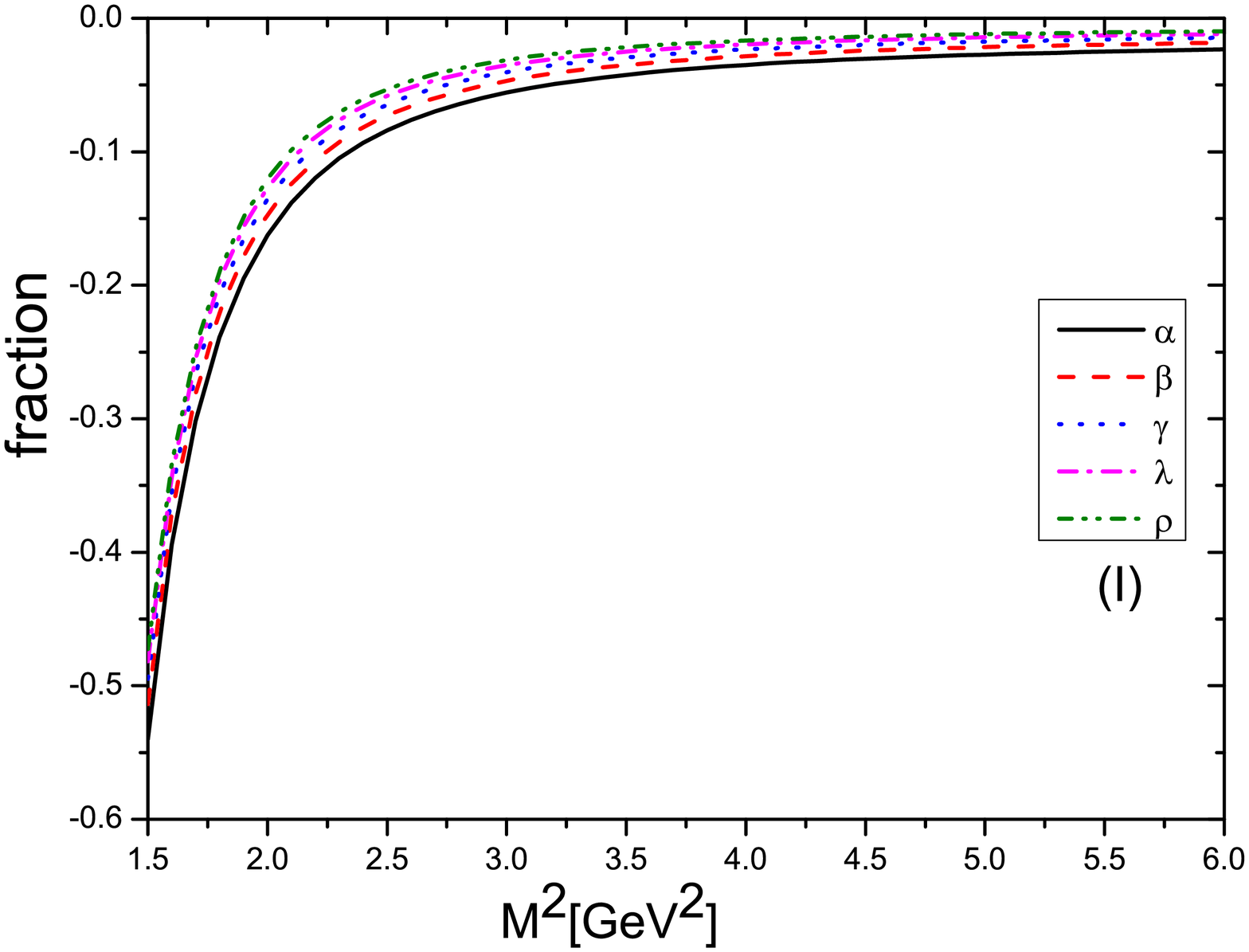}
\includegraphics[totalheight=5cm,width=7cm]{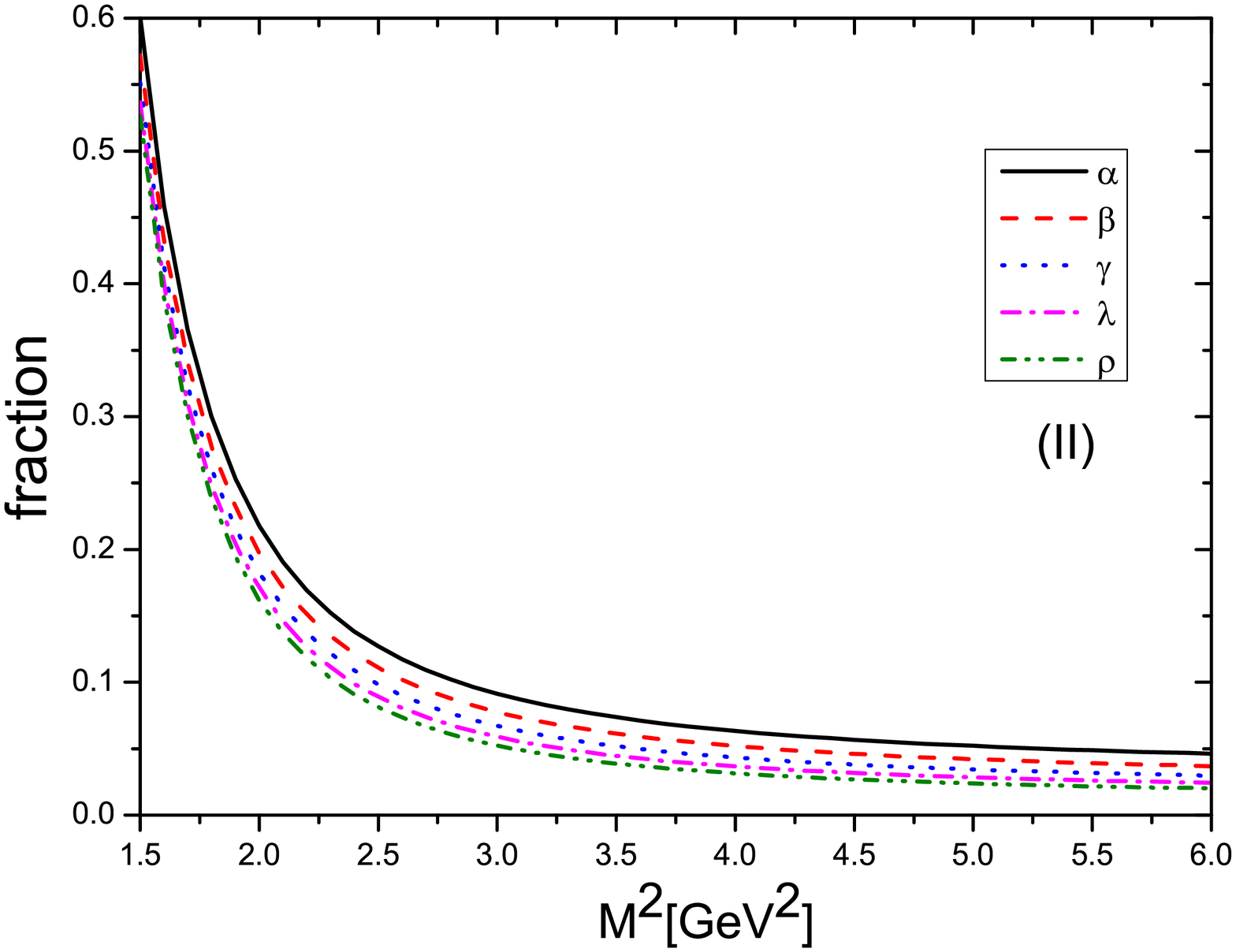}
\caption{The OPE convergence for the $J^{P}=2^{-}$ tetraquark state. The $\,\rm{I}$ and $\,\rm{II}$  correspond to the contributions from the $D=6$ term and the $D=5$ term, respectively. Notations $\alpha$, $\beta$, $\gamma$, $\lambda$ and $\rho$  correspond to threshold parameters $\sqrt{s_0}=4.6\,\rm{GeV}$, $4.7\,\rm{GeV}$, $4.8\,\rm{GeV}$, $4.9\,\rm{GeV}$ and $5.0\,\rm{GeV}$, respectively.}\label{fig1}
\end{figure}

\begin{figure}
\centerline{\epsfysize=6.0truecm\epsfbox{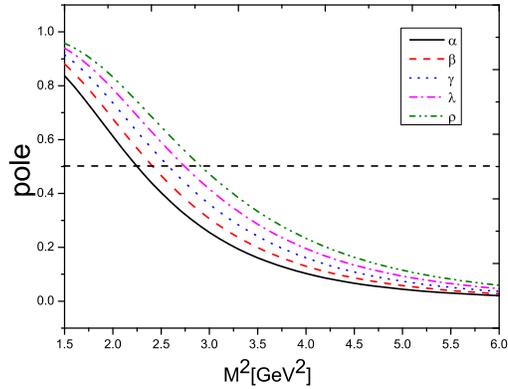}}
\caption{Contributions from pole terms with variation of the Borel parameter $M^2$ in the case of $J^{P}=2^{-}$ tetraquark state. Notations $\alpha$, $\beta$, $\gamma$, $\lambda$ and $\rho$  correspond to threshold parameters $\sqrt{s_0}=4.6\,\rm{GeV}$, $4.7\,\rm{GeV}$, $4.8\,\rm{GeV}$, $4.9\,\rm{GeV}$ and $5.0\,\rm{GeV}$, respectively.}\label{fig2}
\end{figure}

\begin{figure}
\centerline{\epsfysize=6.0truecm
\epsfbox{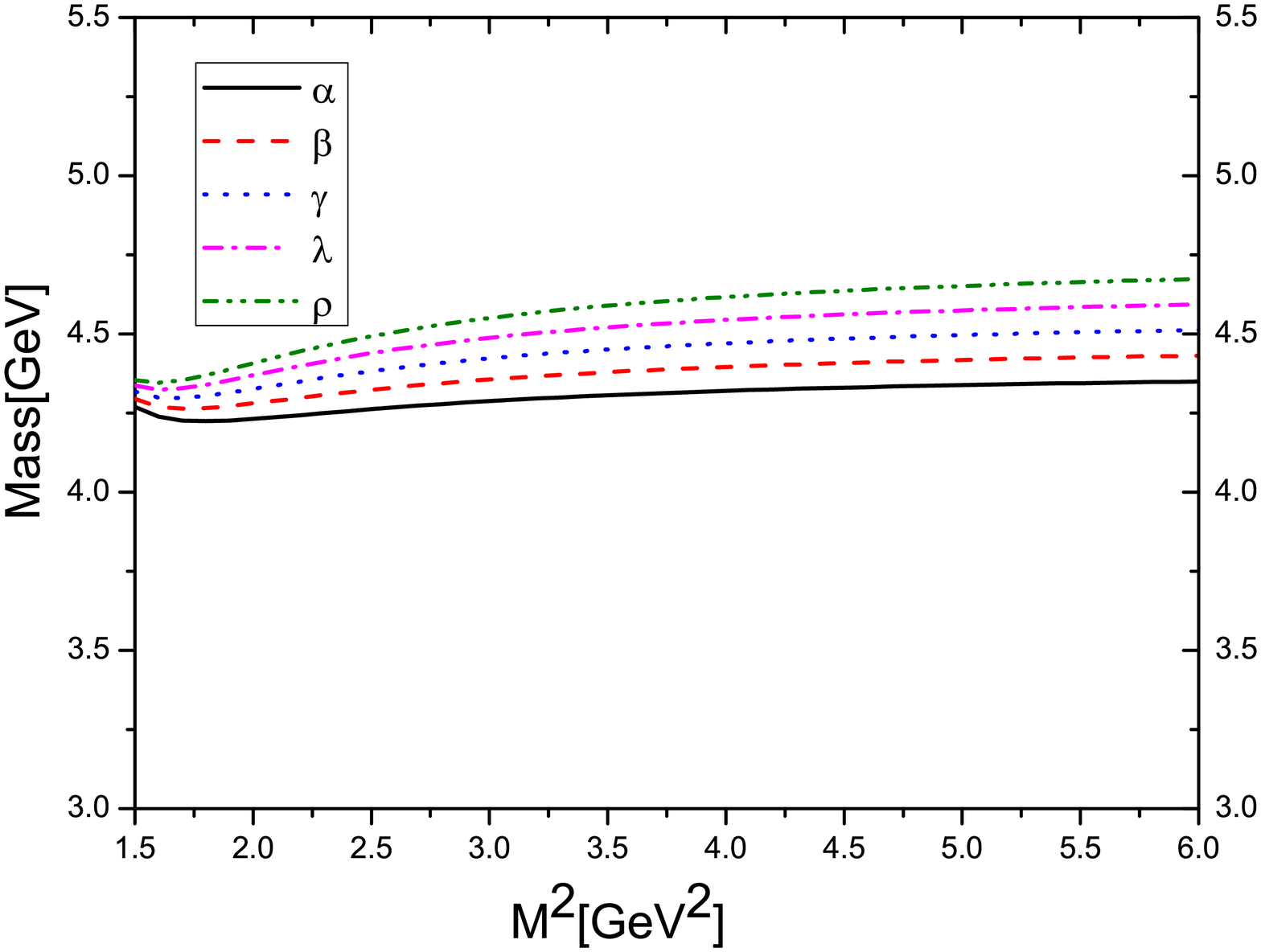}}\caption{
The mass of the $J^{P}=2^{-}$ tetraquark state as
a function of $M^2$. Notations $\alpha$, $\beta$, $\gamma$, $\lambda$ and $\rho$  correspond to threshold parameters $\sqrt{s_0}=4.6\,\rm{GeV}$, $4.7\,\rm{GeV}$, $4.8\,\rm{GeV}$, $4.9\,\rm{GeV}$ and $5.0\,\rm{GeV}$, respectively.}\label{fig3}
\end{figure}

Concretely, the contributions from the high dimension vacuum condensates in the OPE are shown in Fig.\ref{fig1}. We have used $\sqrt{s_0}\geq 4.6\,\mbox{GeV}$. From this figure it can be seen that for $M^2\geq 2.0\,\mbox{GeV}^2$, the contribution of the dimension-$6$ condensate is less than $16\%$ of the total contribution and  the contribution of the dimension-$5$ condensate is less than $20\%$ of the total contribution, which indicate a good Borel convergence. Therefore, we fix the uniform lower value of $M^2$ in the sum rule window as $M^2_{min}= 2.0\,\mbox{GeV}^2$. The upper limit of $M^2$ is determined by imposing that the pole contribution should be larger than continuum contribution. Fig.~\ref{fig2} demonstrates the contributions from the pole terms with variation of the Borel parameter $M^2$. We show in Table~\ref{tab:1} the values of $M^2_{max}$ for several values of $\sqrt{s_0}$. In Fig.\ref{fig3}, we plot the tetraquark state mass in the relevant sum rule window, for different values of $\sqrt{s_0}$. It can be seen that the mass is very stable in the Borel window with the corresponding threshold $\sqrt{s_0}$. The final estimate of the $J^{P}=2^{-}$ tetraquark state is obtained as
\begin{eqnarray}
m_{X}=(4.38\pm0.15)~\mbox{GeV}.
\label{Zmass}
\end{eqnarray}

\begin{table}
\caption{Upper limits in the Borel window obtained from the sum rule for different values of $\sqrt{s_0}$.}\label{tab:1}
\begin{center}
\begin{tabular}{|c|c|}
\hline
$\sqrt{s_0}~(\mbox{GeV})$ & $M^2_{max}(\mbox{GeV}^2)$\\ \hline
4.6 &2.3\\ \hline
4.7 &2.4 \\ \hline
4.8 &2.6 \\ \hline
4.9 &2.8 \\ \hline
5.0 &2.9\\ \hline
\end{tabular}
\end{center}
\end{table}

In summary, by assuming $X(3872)$ as a $[cq][\bar{c}\bar{q}]$ tetraquark state with quantum numbers $J^{P}=2^{-}$, the QCDSR approach has been applied to calculate the mass of the resonance. Our numerical results are $m_{X}=(4.38\pm0.15)~\mbox{GeV}$, which indicates that $X(3872)$ is unlikely to be a $J^{P}=2^{-}$ tetraquark state. Thus, $J^{P}=1^{+}$ assignment for the quantum numbers of the $X(3872)$ is favored.

\begin{acknowledgments}
This work was supported in part by the National Natural Science
Foundation of China under Contracts Nos.10975184, 11047117, 11105222 and 11105223.
\end{acknowledgments}


\begin{thebibliography}{99}
\bibitem{Xobservation} S.-K. Choi {\it et al.} [Belle Collaboration], Phys. Rev. Lett. {\bf 91}, 262001 (2003).
\bibitem{CDF} D. Acosta {\it et al.} [CDF II Collaboration], Phys. Rev. Lett. {\bf 93}, 072001 (2004).
\bibitem{D0} V. M. Abazov {\it et al.} [D$\emptyset$ Collaboration], Phys. Rev. Lett. {\bf 93}, 162002 (2004).
\bibitem{BABAR} B. Aubert {\it et al.} [BABAR Collaboration], Phys. Rev. D {\bf 71}, 071103 (2005).
\bibitem{Belle2} S.-K. Choi {\it et al.} (Belle Collaboration), Phys. Rev. D {\bf 84}, 052004 (2011).
\bibitem{Abulencia:2006ma}
  A.~Abulencia {\it et al.}  [CDF Collaboration],
  Phys. Rev. Lett. {\bf 98}, 132002 (2007).

\bibitem{Abulencia:2005zc}
  A.~Abulencia {\it et al.}  [CDF Collaboration],
  Phys. Rev. Lett. {\bf 96}, 102002 (2006).

\bibitem{Belle-X-gamma-psi}
K.~Abe {\it et al.} [Belle Collaboration], hep-ex/0505037, hep-ex/0505038.

\bibitem{Swanson:2006st}
  E.~S.~Swanson,
  Phys. Rept. {\bf 429}, 243 (2006)

\bibitem{Eichten:2007qx}
  E.~Eichten, S.~Godfrey, H.~Mahlke and J.~L.~Rosner,
  Rev. Mod. Phys. {\bf 80}, 1161 (2008).

\bibitem{Godfrey:2008nc}
  S.~Godfrey and S.~L.~Olsen,
  Ann. Rev. Nucl. Part. Sci. {\bf 58}, 51 (2008).

\bibitem{Voloshin:2007dx}
  M.~B.~Voloshin,
  Prog. Part. Nucl. Phys. {\bf 61}, 455 (2008).

\bibitem{Zhu:2007wz}
  S.~L.~Zhu,
  Int. J. Mod. Phys. E {\bf 17}, 283 (2008).

\bibitem{Klempt:2007cp}
  E.~Klempt and A.~Zaitsev,
  Phys. Rept. {\bf 454}, 1 (2007).

\bibitem{Nora} N.~Brambilla {\it et al.}, Eur. Phys. J. C {\bf 71}, 1534 (2011).

\bibitem{svzsum}M.~A.~Shifman, A.~I.~Vainshtein, and V.~I.~Zakharov, Nucl. Phys. {\bf B147}, 385 (1979); {\bf B147}, 448 (1979); V.~A.~Novikov, M.~A.~Shifman, A.~I.~Vainshtein, and V.~I.~Zakharov, Fortschr. Phys. {\bf 32}, 585 (1984); L.~J.~Reinders, H.~R.~Rubinstein, and S.~Yazaki, Phys. Rep. {\bf 127}, 1 (1985). S.~Narison, QCD Spectral Sum Rules, World Scientific, Singapore, 1989; P.~Colangelo and A.~Khodjamirian, in: M.~Shifman (Ed.), At the Frontier of Particle Physics: Handbook of QCD, vol. 3, Boris Ioffe Festschrift, World Scientific, Sigapore, 2001, pp. 1495-1576, arXiv:0010175; A.~Khodjamirian, talk given at Continuous Advances in QCD 2002/ARKADYFEST, arXiv:0209166.
\bibitem{Nielsenmix} R.~D.~Matheus, F.~S.~Navarra, M.~Nielsen and C.~M.~Zanetti,
  Phys. Rev. D {\bf 80}, 056002 (2009) .
\bibitem{Nielsenradia} F.S. Navarra, M. Nielsen, Phys. Lett. {\bf B639}, 272 (2006).
M.~Nielsen and C.~M.~Zanetti, Phys. Rev. D {\bf 82}, 116002 (2010).

\bibitem{BABAR2} P. del Amo Sanchez {\it et al.} [BABAR Collaboration],
Phys. Rev. D {\bf 82}, 011101(R) (2010).
\bibitem{Yu1007}
Yu Jia, Wen-Long Sang, and Jia Xu
  arXiv:hep-ph/1007.4541
\bibitem{CDF2} B. Aubert {\it et al.} [BABAR Collaboration], Phys. Rev. Lett. {\bf 102}, 13
    2001 (2009).
\bibitem{Belle3} G. Gokhroo {\it et al.} [Belle Collaboration], Phys. Rev. Lett. {\bf 97}, 162002 (2006).
\bibitem{Kalash1008}
Yu. S. Kalashnikova and A. V. Nefediev
Phys. Rev. D {\bf 82}, 097502 (2010).
\bibitem{Chao}
Ying Fan, Jin-Zhao Li, Ce Meng, and Kuang-Ta Chao
  arXiv:hep-ph/1112.3625
\bibitem{Polosa} T. J. Burns, F. Piccinini, A. D. Polosa and C. Sabelli,
 Phys. Rev. {\bf D82}, 074003 (2010).
\bibitem{technique}R.~D.~Matheus, S.~Narison, M.~Nielsen, and J.~M.~Richard, Phys. Rev. D {\bf75}, 014005 (2007); S.~H.~Lee, K.~Morita, and M.~Nielsen, Phys. Rev. D {\bf78}, 076001 (2008); M.~E.~Bracco, S.~H.~Lee, M.~Nielsen, R.~R.~daSilva, Phys. Lett. B {\bf671}, 240 (2009); S.~H.~Lee, K.~Morita, and M.~Nielsen, Nucl. Phys. A {\bf815}, 29 (2009); R.~M.~Albuquerque and M.~Nielsen, Nucl. Phys. A {\bf815}, 53 (2009).
\bibitem{technique1}J.~R.~Zhang and M.~Q.~Huang, Phys. Rev. D {\bf 77}, 094002 (2008); Phys. Rev. D {\bf 78}, 094007 (2008); Phys. Rev. D {\bf 78}, 094015 (2008); Phys. Lett. B {\bf 674}, 28 (2009).
\bibitem{PDG}C.~Amsler {\it et al.}, (Particle Data Group), Phys. Lett. B {\bf 667}, 1 (2008).
\end{thebibliography}
\end{document}